\documentclass[prb,aps,twocolumn,showpacs,preprintnumbers,floats,amsmath,amssymb]{revtex4}

\usepackage{graphicx}
\usepackage{dcolumn}
\usepackage{bm}

\def \be{\begin{equation}}
\def \ee{\end{equation}}

\def\moy#1{\left\langle #1 \right\rangle}

\begin{document}

\title{Anatomy of the quantum melting of the two dimensional Wigner crystal}

\author{X. Waintal}
\affiliation{Nanoelectronics group, Service de Physique de l'\'Etat Condens{\'e},
CEA Saclay, 
91191 Gif-sur-Yvette cedex, France }

\date{\today}

\begin{abstract}
The Fermi liquid-Wigner crystal transition in a two dimensional electronic
system is revisited with a focus on the nature of the fixed node approximation 
done in quantum Monte Carlo calculations. 
Recently, we proposed (Phys. Rev.  Lett. {\bf 94}, 046801 (2005) ) 
that for intermediate densities, a hybrid phase 
(with the symmetry of the crystal but otherwise liquid like properties) is more stable than both
the liquid and the crystal phase. Here we confirm this result both in the thermodynamic and
continuum limit. The liquid-hybrid transition takes place at $r_s^*=31.5\pm 0.5$. We find that the
stability of the hybrid phase with respect to the crystal one is tightly linked to its delocalized 
nature. We discuss the implications of our results for various transition scenarii (quantum hexatic phase,
supersolid, multiple exchange, microemulsions) proposed in the literature.
\end{abstract}

\maketitle

\section{Introduction}

\begin{figure}
\vglue +0.05cm
\includegraphics[width=8cm]{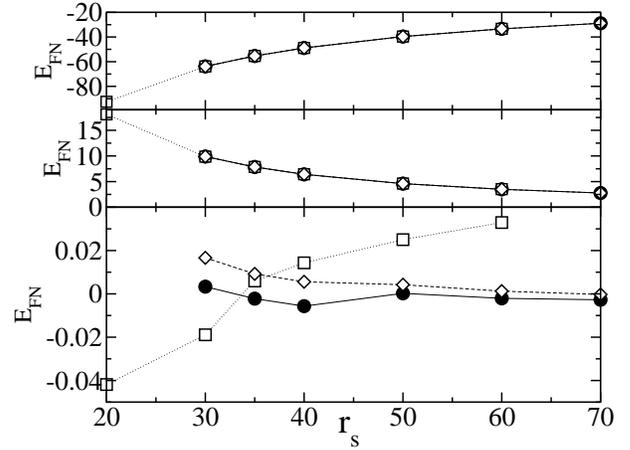}
\caption{\label{fig:intro} Quantum Monte-Carlo energy as a function of $r_s$ for a system of
$56$ electrons in $210\times 208$ sites for the liquid phase (squares), solid phase (diamonds) and
hybrid phase (full circles).  Upper panel: total energy, the three phases are indistinguishable. Middle panel:
total energy minus the Madelung (electrostatic) energy $-c_1/r_s$ with $c_1=2.2122$. 
Lower panel: total energy minus the Madelung energy and the energy of the crystal's 
phonons $c_{1/2} r_s^{-3/2}$ with $c_{1/2}=1.624$. The energies are given in mRy. For holes in a GaAs
heterostructure, a Rydberg  corresponds roughly to 350 Kelvin ($\epsilon\approx 13\epsilon_0$, 
$m^*\approx 0.38 m_e$).}
\end{figure}

The competition between electrostatic and kinetic energy in a two dimensional electron
gas is a problem which is simple to formulate (what is the phase diagram of $N$ electrons
on a surface $S$ at zero temperature) yet difficult to tackle. 
One reason for this difficulty lies in the smallness
of the difference in energy between different phases (The Fermi liquid which is stable at high
density when the kinetic energy dominates~\cite{pines1966} 
and the electrostatically favored (Wigner) triangular crystal, 
stable at low density~\cite{wigner1934}). 
 In the region of interest in this study, $20\le r_s \le 80$ 
(the dimensionless parameter $r_s\propto\sqrt{S/N}$ controls the ratio of the electrostatic energy
over the kinetic energy) where there is a strong competition
between the two kinds of energy, the difference of energy between the two phases 
is only of the order of $0.1\%$ of the total energy so that
very accurate methods were needed to study this problem. 
To illustrate this point, we plot in the Fig.~\ref{fig:intro} the energy of these two phases
as a function of $r_s=m^* e^2/(\hbar^2\epsilon \sqrt{\pi n})$ 
($e$ being the electronic charge, $\epsilon$ the dielectric constant, $m^*$ the
effective mass and $n=N/S$ the electronic density). 
At large distances, the physics is entirely controlled by
electrostatic, and important amount of energy are involved. Once we add a positive background
to the system and make it globally neutral, we arrive at the energies plotted 
in the upper panel of Fig.~\ref{fig:intro}: on the scale of the plot the liquid and crystal are 
completely indistinguishable. If we remove the Madelung energy (excess electrostatic energy of the
uniform background with respect to the crystal), we still cannot distinguish between the two phases 
(middle panel of Fig.~\ref{fig:intro}). It is only
after we have removed the zero point fluctuation energy of the crystal that the curves become different 
and show a crossing at $r_s\approx 37$ where the Wigner crystallization has been
 believed to occur~\cite{imada1984,tanatar1989} (lower panel of Fig.~\ref{fig:intro}).

In a seminal article in 1989, Tanatar and Ceperley~\cite{tanatar1989} 
used a fixed node quantum Monte Carlo (FN-QMC) technique~\cite{foulkes2001} to locate
the critical value $r_s\approx 37\pm 5$ where the quantum melting of the Wigner crystal occurs.
 Their work was followed by more precise numerics~\cite{rapisarda1996} and a better
description of the liquid phase~\cite{attaccalite2002,kwon1993} that included backflow corrections.
On the other hand there were indications that this
scenario of a (first order) direct transition between the Wigner crystal and the Fermi liquid phase 
might miss some of the physics. For instance, the quantum melting of the bosonic Wigner crystal~\cite{palo2004,bernu2003}
is found (by similar QMC calculations) to occur at $r_s\approx 60$ so that for $37\le r_s\le 60$
the fermionic statistics (from which all the difficulties of the QMC calculations come, as we shall see)
is crucial to stabilize the crystal.  Also, the classical melting 
\cite{andrei1988,kosterlitz1973,platzman1974,halperin1978} as a function of temperature
does not occur in a one step process.
The  system first looses its translational order but retains some orientational order (hexatic
phase~\cite{halperin1978}) while at a higher temperature, any order disappears. The possibility of
an intermediate quantum hexatic phase was put forward in~\cite{oganesyan2001,barci2003}.
In fact, it was recently argued on rather general grounds that a direct first order transition
is simply impossible in this system~\cite{jamei2005,spivak2004} but should occur through a 
serie of intermediate phases (including bubbles or stripes of one phase in the other). 
The possibility of a supersolid phase 
analogous to the one proposed by Andreev and Lifshitz~\cite{andreev1969} 
has also been considered~\cite{spivak2003, katomeris2003,nemeth2003}. All these
indications~\cite{nagy1999} lead us recently to revisit the melting of the Wigner crystal in the framework of the 
FN-QMC calculations. We introduced a third phase called hybrid phase (as it had the symmetry of the 
crystal but is otherwise delocalized) and showed that it was more stable than both the liquid 
and crystal phase (Full circles in the lower panel of Fig.~\ref{fig:intro}) in the intermediate 
density range~\cite{falakshahi2005} giving a strong support to the more exotic scenarios. 

To understand the status of our result however, it is of prime importance to understand the nature of the
(fixed node, FN) approximation involved in the FN-QMC technique. In practice, 
the FN-QMC algorithm is fed with a wave-function called the {\it guiding wave function} 
(GWF) that has to be given explicitly, and that should be as close as possible to the 
ground state of the system. The FN-QMC algorithm projects the GWF onto  
the (true) ground state of the system. However, to avoid the notorious sign problem
that arises in QMC simulations involving fermions, one approximation is introduced: the 
projection is done with the constraint that {\it the sign of the wave function (nodal structure)
remains unchanged} at every point of the Hilbert space. The method
gives the best wave function for a given structure of the nodes 
of the GWF and is in this sense variational~\cite{haaf1995}. The physical meaning of the FN approximation is
not completely obvious. Implicit in the interpretation given above of the FN-QMC results is that a GWF is 
associated to a phase (crystal, liquid or hybrid) the stability of which can therefore be studied. However,
this paradigm should not be taken too literally and will be revisited in this article.

After presenting our model and the FN-QMC method in section II, we perform FN-QMC calculations on a small
system of $4$ electrons (section III). As this system can be studied exactly, it is an interesting tool
to study the nature of the FN approximation. We find that in addition to the FN results at large imaginary
(projection) time, important information is embedded in the evolution of the results from the variational
calculation to the full FN-QMC results. In section IV, we explain the construction of the hybrid GWF, and
show that its nodal structure corresponds to delocalized waves. Section V is devoted to a characterization 
of the physics associated with the hybrid GWF. This is done through a systematic study 
of various physical quantities, and in particular
of their evolution between the variational and FN-QMC calculations. We find that the success of the hybrid GWF
with respect to the crystal one is closely linked to its delocalized nature.
Section VI contains a detailed discussion
of various technical aspects (thermodynamic limit, lattice effects, mixed and unbiased estimates,...) 
The chief result of section VI is a precise determination of the critical value  $r_s^*=31.5\pm 0.5$.
at which the liquid-hybrid transition takes place (in both the continuum and thermodynamic limit).
 In the discussion 
section VII, we discuss the implications of our results to the scenarii proposed in the literature.

\section{Model and Method.}

\subsection{System under consideration}
The system under consideration consists of $N$ spinless electrons on a rectangular $L_x\times L_y$ grid 
with nearest neighbor hopping and long range Coulomb repulsion. To avoid strong finite size effects on
the electrostatic energy, the system is repeated periodically and fills the whole two dimensional plane.
In practice, we use periodic boundary conditions for the hopping terms and the effective two-body interaction
is obtained from the bare Coulomb interaction using the Ewald summation technique. The system Hamiltonian 
reads~\cite{trivedi1990}
\be
\label{eq:model}
H=-t\sum_{\langle\vec r,\vec r'\rangle}c_{\vec r}^\dagger c_{\vec r'}
+\frac{U}{2} \sum_{\vec r\ne\vec r'} V(\vec r-\vec r') n_{\vec r} n_{\vec r'}  + \lambda.
\ee
where the operator $c_{\vec r}^\dagger$ ($c_{\vec r}$) creates (destroys) an electron on
point $\vec r$ with the standard anticommutation relation rules, the sum $\sum_{\langle\vec r,\vec r'\rangle}$ 
is done on the nearest neighbor points on the grid and $t$ is the corresponding hopping amplitude. 
The density operator reads $n_{\vec r}=c_{\vec r}^\dagger c_{\vec r}$.  
$U$ is the effective strength of the two body interaction $V(\vec r)$ which reads,
\be
V(\vec r) = \sum_{\vec L} \frac{1}{|\vec r +\vec L|} {\rm Erfc}(k_c |\vec r +\vec L|)
\ee
$$+\frac{2\pi}{L_x L_y} \sum_{\vec K\ne \vec 0} 
\frac{1}{|\vec K|} 
{\rm Erfc}(|\vec K|/(2 k_c))\cos (\vec K\cdot\vec r) .
$$
In the previous equation, $k_c$ is a (irrelevant) cut off. The vector $\vec L$
takes discrete values $\vec L= (n_x L_x,n_y L_y)$ with $n_x$ and $n_y$ integer numbers.
The vector $\vec K$ also takes discrete values, $\vec K= (\frac{2 \pi}{L_x} n_x,
\frac{2 \pi}{L_y} n_y)$ and $(n_x,n_y)\ne (0,0)$. The complementary error function is
defined as ${\rm Erfc}(r)=\frac{2}{\sqrt{\pi}} \int_x^\infty e^{-t^2} dt$.
In order to assure electrostatic neutrality we add a positive continuum background 
\be
\lambda/N = 4 t + U \tilde V(\vec 0) -2 U \nu \sqrt{\pi}/k_c - 2 U k_c/\sqrt{\pi}
\ee
where $\nu=\frac{N}{L_x L_y}$ is the average number of electrons per site and
$\tilde V(\vec r)=V(\vec r)$ with the restriction that the sum over $\vec L$
does not includes the null vector.

The presence of the grid can be understood either as a discretization of the continuum problem
(and the nearest neighbor hopping corresponds to the discretized Laplacian) or as a tight-binding 
approach to two dimensional electron systems where each site corresponds to an atomic orbital. 
In conventional 2d systems (say GaAs/GaAlAs heterostructures), 
densities range from $n_s\approx 10^{12}{\rm cm}^{-2}$ down 
to $n_s\approx 10^{10}{\rm cm}^{-2}$. Since the distance between nearest atoms is of the 
order of a few Angstroms, this would lead to $\nu\sim 10^{-4}$ to $\nu\sim 10^{-2}$ electrons per 
site in our tight-biding picture. In this study,
we will study systems with $\nu=1/56$, $\nu=1/224$ and $\nu=1/780$ close to realistic values.

The presence of the
jellium (which is merely a constant term and thus cannot affect the physics) allows us to make a
quantitative contact with the literature in the continuum limit as $\nu\rightarrow 0$. 
In this limit, the physics depends only on the $r_s$ parameter which reads, $r_s=U/(2 t \sqrt{\pi \nu})$ while 
the Rydberg unit of energy is $Ry=U^2/(4t)$. In the following, unless specifically
stated, we shall measure all energies $E$ in unit of $2\pi N\nu t$ (i.e. energy per particle in unit
of the Fermi energy of the non-interacting problem.) With this normalization, 
the energy of the system at $r_s=0$ in the thermodynamic $N\gg 1$ and continuum $\nu\ll 1$ limit
is $E=1$ so that results in Rydberg can be obtained by multiplying the energies 
(in unit of $2\pi N\nu t$) by  $2/r_s^2$.

 We have added a very small disorder $\sum_{\vec r} v_{\vec r}n_{\vec r}$ in order to 
lift the degeneracies of the non interacting problem.
$v_{\vec r}$ are independent and uniformly distributed inside $[-W/2,+W/2]$. We choose $W=10^{-3}$
corresponding to an extremely large ratio $l/\lambda_F= 96\nu t^2/W^2 \sim 10^6$ of the mean free 
path $l$ over Fermi wave length $\lambda_F$. We explicitly checked that our result are insensitive 
to the presence of this disorder.

The grid has been chosen to accommodate a (almost) triangular Wigner crystal without distortion. Hence
we use systems of $N=Q\times R$ electrons ($R$ ($R$ even) lines of $Q$ particles) in
$Q d_x \times R d_y$ sites. The triangular crystal requires $d_y/d_x=\sqrt{3}/2\approx 0.866$. 
In this work, we focus on  $d_y/d_x=14/16\approx 0.875$ and $d_y/d_x=26/30\approx 0.866$ 
for which we find that the distortion is negligible.

\subsection{The Green Function Monte-Carlo method in the Fixed Node approximation}
We aim to sample the ground state $|\Psi_0\rangle$ of $H$ and its 
corresponding ground state energy $E_0$ with the Green Function Monte-Carlo technique.
The idea behind this method is to project an initial (variational)
guiding wave-function $|\Psi_{G}\rangle$ on the exact ground state $|\Psi_0\rangle$  
by applying the operator $e^{-H \beta}$ in a stochastic way. In practice, one works in the 
many particle basis $R=( \vec r_1,\vec r_2 ...\vec r_N)$ and applies the
function 
\be
\label{eq:green}
G_{R'R}= \delta_{R'R} - \tau \Psi_{G}(R')\left[ H_{R'R} - \omega \delta_{R'R} \right] \Psi^{-1}_{G}(R)
\ee 
where $\Psi_{G}(R)=\langle R |\Psi_{G}\rangle$, $\tau$ is a (small) time step and
$\omega$ is an unimportant offset of the energies (that should be roughly set to the ground state energy of the
system, here we took $\omega/(2\pi\nu t)= 1-r_s$). In the absence of the $\Psi_{G}(R)$, $G$ is a
discretized version of the operator $e^{-H \beta}$ for a small (imaginary) time step $\tau$. Upon applying $n=\beta/\tau$ times $G$ on a vector $\Psi^2_{G}(R)$, the $\Psi_{G}(R')$ on the left of Eq.(\ref{eq:green}) are 
canceled by the $\Psi^{-1}_{G}(R)$ on the right so that one obtains 
\be
I_n=\sum_{R'R} (G^n)_{R'R} \Psi^2_{G}(R)= \langle\Psi_{G}|e^{-(H-\omega) n\tau}|\Psi_{G}\rangle
\ee 
from which one can extract the ground state energy (as $I_{n+1}=e^{-(E_0-\omega)\tau} I_n$ 
when $\beta=n\tau\rightarrow\infty$).

The stochastic implementation of this scheme is based on the Green Function Monte Carlo 
for lattice Hamiltonians introduced in Ref.~\onlinecite{trivedi1990} to which we refer for more details. 
The algorithm to update the Slater determinants used in the calculation of
$\Psi_{G}(R)/\Psi_{G}(R')$ can be found in Ref.~\onlinecite{ceperley1997}. By sampling directly the time 
spent by the walkers at one point of the Hilbert space using
the algorithm described in Ref.\onlinecite{trivedi1990} we can use arbitrary small time steps $\tau$
without any loss in computing time and hence effectively work in continuum (imaginary) time. 
Instead of using the standard branching technique, the control of the walkers population is done
using a fixed number of walkers and the reconfiguration algorithm introduced by 
Sorella~\cite{sorella1998}. This algorithm allows to avoid the bias introduced 
in the branching technique by artificially controlling the walker population. 
Quantum averages of physical quantities $\langle\dots\rangle$ are calculated using the
forward walking technique~\cite{sorella1998}, and hence do not suffer from the bias of mixed
estimates.

So far the scheme is essentially exact. However, it suffers from the usual 
``sign problem'', the sign of $G_{R'R}$ fluctuates so that the statistical accuracy 
decreases exponentially with $\beta$, and it is of
little practical use. One way out of the sign problem is the Fixed-Node approximation where
one forbids the sign of the wave function to change upon applying $G$ (hence the name: the nodal
surface where the wave-function changes of sign remains the same than the one of $\Psi_{G}(R))$. 
The practical implementation of the fixed-node approximation on a grid is done\cite{haaf1995} by
replacing $H$ by an effective Hamiltonian $H^{\rm FN}$ that depends on the GWF. 
$H^{\rm FN}_{R'R}$ is equal to $H_{R'R}$ when $G_{R'R}>0$. When $G_{R'R}<0$, the link is
cut $H^{\rm FN}_{R'R}=0$ and is replaced by an effective potential  
$H^{\rm FN}_{RR}= H_{RR}+\sum_{R'} \theta(-G_{R'R}) \Psi_{G}(R') H_{R'R}\Psi^{-1}_{G}(R)$ where
$\theta(x)$ is the Heaviside function (this corresponds to replace the value of the wave-function
on those sites $R'$ by the surmise $\Psi_{G}(R')$). The fixed node approximation can be thought as
a ``supervariational'' technique where the amplitude of the wave-function is optimized at every point
of the Hilbert space while its sign remain fixed. It can be proved indeed that the energies $E_{\rm FN}$
calculated with $H_{\rm FN}$ are larger than the true ground state energy $E_0$ but smaller than 
the variational energy associated with the guiding wave-function~\cite{haaf1995}. 
The fixed node approximation in a
lattice looks {\it a priori } more drastic than in the continuum since we do impose the ratio of the
wave-function across the nodal surface. However, for $\nu\ll 1$ the fraction of 
``nodal'' sites goes to zero and the technique becomes equivalent to the fixed node 
diffusive Monte-Carlo used in the continuum~\cite{tanatar1989}.

\begin{figure}
\vglue +0.05cm
\includegraphics[width=8cm]{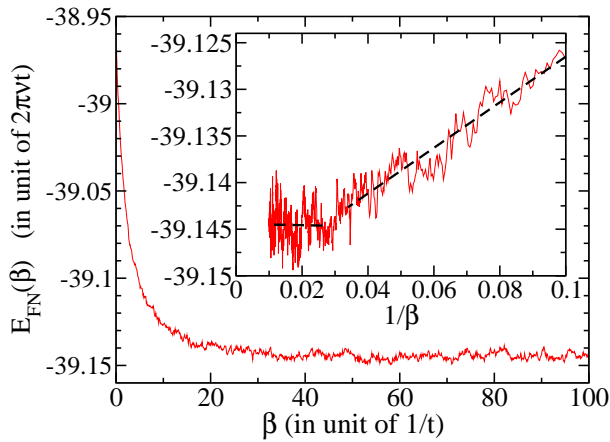}
\caption{\label{fig:rawdata}(color online).
Typical curve of $E_{\rm FN}(\beta)$ as a function of $\beta$. System of 32 particles in $64\times 112$ sites at $r_s=40$ for a liquid GWF with $A=4.5$. Here,
 $8.10^5$ walkers were used, $\omega=-39$ and $\tau=10^{-5}/t$. Inset: same curve as a function
of $1/\beta$. The dashed lines are linear fits to guide the eye. }
\end{figure}

As an illustration, we present in Fig.~\ref{fig:rawdata} a typical trace obtained for 32 particles at $r_s=40$. 
$E_{\rm FN}(\beta=0)$ corresponds to the variational energy 
$\langle\Psi_{G}|H|\Psi_{G}\rangle/\langle\Psi_{G}|\Psi_{G}\rangle$. After an initial rapid decrease, the FN energy  (per particle) $E_{\rm FN}(\beta)$ decreases
slowly as $1/\beta$ (see inset) and then saturates above an imaginary time $\beta_{\rm sat}\propto 1/(\nu t)$.
The energy is estimated by further averaging the result over $\beta$ for $\beta>\beta_{\rm sat}$. The typical
achieved accuracy  allows us to determine $E_{\rm FN}(\beta)$ with a precision better than $\pm 0.001$. This 
remarquable precision (here the relative accuracy is $\sim 10^{-5}$) should be contrasted with the fact that
important changes in the physics can lead to very tiny changes in energy. For instance the condensation energy
in a superconductor is only a very small fraction $\sim (\Delta/E_F)^2 \sim 10^{-6}$ ($\Delta$ superconducting gap,
$E_F$ Fermi energy) of the total energy (only electrons near the Fermi surface form Cooper pairs). The formation 
of a Wigner crystal (localized in real space) from a Fermi liquid (localized in momentum space) corresponds
however to a complete reorganization of the system  and the changes in energy, though small $\sim 0.01$, can
be measured with the GFMC technique.   
\subsection{Guiding wave functions.} 
Central in the Fixed-Node technique is the choice of the GWFs used in the calculations.
Here, the guiding wave-functions have the general form of a Slater determinant multiplied by 
a Jastrow function,
\be
\Psi_G(R)=
{\rm Det\ }[\phi_i(\vec r_j)]\times \prod_{i<j} J(| \vec r_i - \vec r_j|). 
\ee
The Jastrow part takes Coulomb interaction into account  
by introducing correlations between electrons. Since it 
has no nodes, the FN results should not depend on its particular
form (as we explicitly checked). We use modified Yukawa functions\cite{stevens1973},
\be
J(r)=\exp \left[\frac{a A(r_s)}{r}(1-e^{-B(r_s) r/a})\right],
\ee 
where $a=1/\sqrt{\pi\nu}$ is the average distance between electrons.
$A(r_s)$ and $B(r_s)$ are (optimized) variational parameters. To avoid the
Coulomb singularity when two electrons get close to each other, we impose
the cusp condition~\cite{foulkes2001} that reads $B=\sqrt{r_s/A}$ for the modified Yukawa.

The Slater determinant of one-body wave functions, ${\rm Det\ }[\phi_i(\vec r_j)]$
enforces the antisymmetric nature of the fermionic wave function and is 
responsible for the nodal structure of the GWF. The GWF used in the literature for
the study of the melting of the Wigner crystal~\cite{tanatar1989,rapisarda1996} 
are of two kind, adapted to the two limits
 of very low (large) $r_s$:
\begin{itemize}
\item For the liquid state the GWF $\Psi_{\rm liq}$ is constructed out of 
plane waves $\phi_i(\vec r_j)\propto e^{i\vec k_i \cdot \vec r_j}$ with a well
defined Fermi surface. $\Psi_{\rm liq}$ is the exact ground state at $r_s=0$.
\item For the crystal GWF $\Psi_{\rm cry}$, localized orbitals
$\phi_i(\vec r_j)\propto e^{-(\vec r_j-\vec u_i)^2/(2 d_0^2)}$ are used.
Here the $\vec u_i$ with $i\in\{1\dots N\}$ stand for the positions of the electrons
in the classical crystal. 
$\Psi_{\rm cry}$ provides the exact ground state in the continuum at large $r_s$
with the variational parameter $d_0\propto a/r^{1/4}_s$. This GWF captures the two leading terms
of the large $r_s$ expansion of the energy (Madelung and zero point fluctuation energy) that reads,
\be
E = -c_1 r_s/2 +c_{1/2} \sqrt{r_s}/2 + \cdots
\ee
with $c_1=2.2122$ and $c_{1/2}=1.624$.
\end{itemize}
In this article, we will use the two previously mentioned GWF, and introduce a third (hybrid)
one that somehow interpolates between the crystal (real space) and the liquid (momentum space).
 
\section{The sign problem and the Fixed Node approximation}

\begin{figure}
\vglue +0.05cm
\includegraphics[width=6cm]{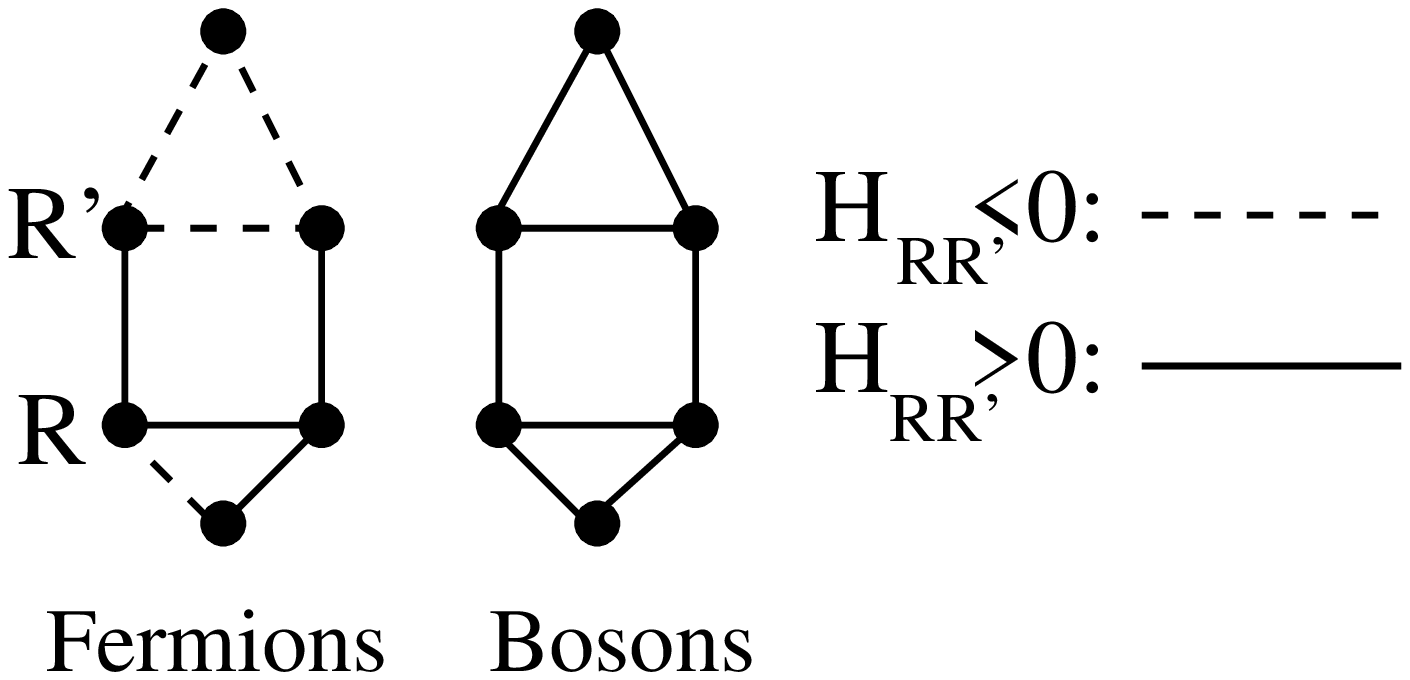}
\caption{\label{fig:signproblem}Schematic view of Eq.(\ref{eq:schro}). The circles symbolize sites $R$ in
Hilbert space and the full (dashed) lines positive (negative) off-diagonal matrix elements $H_{RR'}$. The 
corresponding lattice is frustrated for fermions.}
\end{figure}

In this section, we come back to the sign problem, and investigate the nature of the approximation
involved in the Fixed Node approximation. One way to understand the sign problem is to consider it as a
frustration problem in Hilbert space: we seek to solve the Schr\"odinger equation which formally reads
in $R$ space,
\be
\label{eq:schro}
\sum_{R'} H_{RR'} \Psi_0(R') = E_0\Psi_0(R).
\ee
Two-body interactions and external potentials are diagonal in the $R$ space and $H_{RR}$ can always
be considered to be positive (by shifting $H$ by a constant). The off-diagonal elements  $H_{RR'}$
come from kinetic energy. They are positive for bosons but are alternatively positive and
negative for fermions as shown schematically in Fig.~\ref{fig:signproblem}. The bosonic case does 
not suffer from any sign problem and $\Psi^{\rm boson}_0(R)$ can be found efficiently. We now consider 
a fictitious Ising model where an Ising variable $s(R)=\pm 1$ is placed on each site $R$ of the Hilbert 
space and is coupled to its neighbors by a coupling $H_{RR'}$ (which can be ferro or antiferromagnetic). 
If this fictitious model had no frustration, then its ground state $s_0(R)=\pm 1$ could be trivially found 
and one can easily verify that  $\Psi^{\rm fermion}_0(R)=s_0(R)\Psi^{\rm boson}_0(R)$ 
would be the exact ground state of the fermionic problem. 
Hence the sign problem arises from the frustrated nature of Eq.(\ref{eq:schro}). From this point of view,
the Fixed-Node approximation consists in cutting some links of this fictitious Ising model so that it is no
longer frustrated and admits $s_0 (R)={\rm sgn} (\Psi_G(R))$ as its ground state. The FN approximation does
not have a simple physical meaning and thus seems difficult to control.

\begin{figure}
\vglue +0.05cm
\includegraphics[width=8cm]{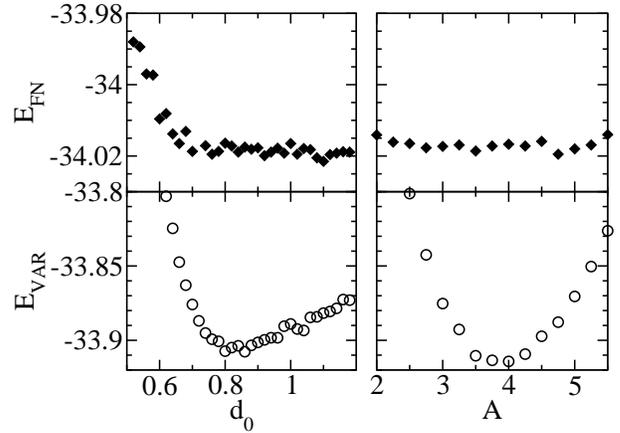}
\caption{\label{fig:var_vs_fn} Comparison between the Fixed Node $E_{\rm FN}$ (upper panels)
and variational $E_{\rm VAR}$ (lower panels) energies as a function of $d_0$ (left panels)
and $A$ (right panel) for a system of $30$ particles in $40\times 42$ sites at $r_s=35$
using the $\Psi_{\rm cry}$ GWF. Left panels: $A=4.5$. Right panels: $d_0=0.76$.}
\end{figure}

In Fig.~\ref{fig:var_vs_fn}, we compare the FN results with the variational (i.e. $\beta=0$)
results for a system of $30$ particles at $r_s=35$. The calculations have been done for the
 $\Psi_{\rm cry}$ GWF so that we can compare the influence of the two parameters $d_0$ and $A$.
While $d_0$ enters in the definition of the Slater determinant and hence of the nodal
structure, $A$ defines the Jastrow part which does not change sign and (in the limit $\nu\ll 1$)
the FN results should thus be insensitive to the choice of $A$. On the right panel of Fig.~\ref{fig:var_vs_fn},
we see that it is indeed the case. The function $E_{\rm FN}(A)$ is flat at the precision of the calculation ($\pm 0.002$
here) while the variational energy $E_{\rm VAR}(A)$ shows a pronounced minimum at $A\approx 4$. On the left panels
of  Fig.~\ref{fig:var_vs_fn}, we plot $E_{\rm FN}$ and $E_{\rm VAR}$ as a function of $d_0$. One would
expect the two curves to present a minimum at the same value of $d_0$ (which is what is usually assumed,
the FN calculations being done in most cases with parameters optimized at the variational level).
However, one observes that for $d_0\ge0.8$ (variational minimum) the FN energy is flat upon increasing
$d_0$. Hence, the nature of the approximation done in a fixed node 
calculation is not simply related to the original variational wave-function used for fixing the 
nodal surface, and the method can potentially capture more physics than originally thought.

To get further insight, we now discuss the instructive example of $4$ electrons in a $6\times 8$ grid.
This system is small enough (the size of the Hilbert space is $\sim 2.10^5$) so that its 
ground state can be found exactly using the Lancsos algorithm and compared with the result of a
Fixed Node calculation. We introduce the density $\rho(\vec r)$ and density-density correlation function
$g(\vec r)$
\be
\rho(\vec r)=\frac{1}{N}\moy{c_{\vec r}^\dagger c_{\vec r}}
\ee
\be
g(\vec r)=\frac{L_xL_y}{N(N-1)}\sum_{\vec h}\moy{c_{\vec r+\vec h}^\dagger c_{\vec h}^\dagger c_{\vec h} c_{\vec r+\vec h}}
\ee
which measures respectively the average number of electrons on site $\vec r$ and the probability to
find an electron on site $\vec r$ knowing 
that there is one particle on site $\vec 0$ (both are normalized to $1$).
From these two quantities, we define the participation ratio 
\be
\rho_2 = 1/\left[L_xL_y \sum_{\vec r} \rho^2(\vec r)\right]
\ee
which roughly measures the number of occupied sites in the ground states ($\rho_2$ interpolates from $1$
to $\nu$ as $r_s$ goes from $0$ to $\infty$) and the ``rigidity'' parameter $g_2$
\be
g_2=\frac{1}{L_xL_y} \sum_{\vec r} \left[g(\vec r)-1\right]^2
\ee
The comparison between the exact result and the FN calculation is presented in Fig.~\ref{fig:exact_vs_qmc}.
In the inset of Fig.~\ref{fig:exact_vs_qmc}, we show the exact $g(\vec r)$ at $r_s=20$. $g(\vec r)$ 
has three well defined maxima at the positions of the classical Wigner crystal. Upon increasing $r_s$, the 
system gets more and more crystal like and these maxima gets more pronounced which leads to an increase 
of $g_2$ (lower panel) and a decrease of $\rho_2$ (middle panel). Figure ~\ref{fig:exact_vs_qmc} shows two
regimes. For $r_s\le 30$, the error on the energy (upper panel) increases linearly while the two other physical
quantities $\rho_2$ and $g_2$ are in very good agreement with the exact result. Hence the FN result, though
calculated with a liquid GWF made of plane waves, is well able to reproduce the ground state of a system
which (at $r_s=30$) is really crystal like. Above $r_s\ge 30$ the error saturates to $0.007$, 
$\rho_2$ is still in
good agreement with the exact result, but the error on $g_2$ starts to increase significantly. At $r_s=70$,
the error on the energy is only $0.01\%$ of the total energy ($E_0\approx 1- r_s$) while the error on
$g_2$ is much larger $\sim 33\%$.

Hence the FN approximation provides a very good description 
of the ground state up to a rather large value of $r_s$ (here around 30) even though the system 
is very crystal like and we started from a liquid GWF. It also provides a good quantitative value
of the energy even for higher values of $r_s$ where the FN approximation no longer describes
accurately the physics. Maybe even more importantly, 
Fig.~\ref{fig:exact_vs_qmc} gives us a clue on how to control the FN approximation:
\begin{figure}
\vglue +0.05cm
\includegraphics[width=8cm]{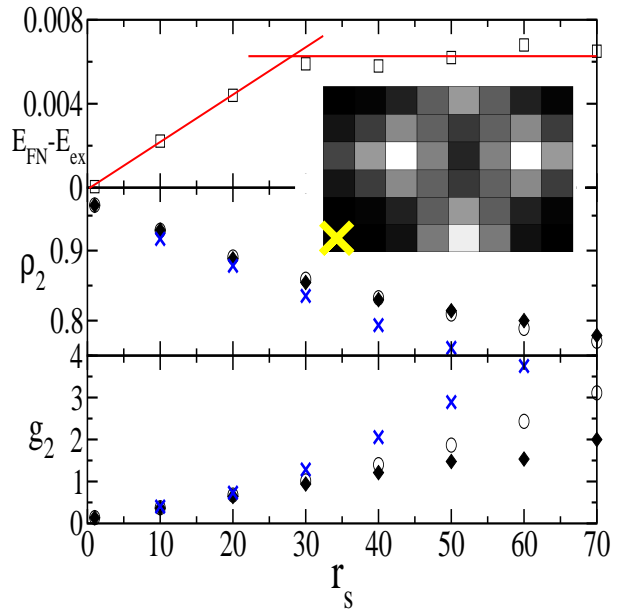}
\caption{\label{fig:exact_vs_qmc}(color online). Comparaison between the Fixed Node
calculation with a liquid GWF and the exact result for a system of $4$ particles in $8\times 6$ sites.
Upper panel: difference between the FN and exact energy. The lines are guide to the eye. Middle panel:
participation ratio  per site $\rho_2$ for the exact (empty circles), FN (full diamonds) and 
variational (crosses) calculation.
Lower panel: idem for the rigidity parameter $g_2$. Inset: Exact density-density correlation function $g(\vec r)$ 
at $r_s=20$. $g(\vec r)$ measures the probability to find a particle on site $\vec r$ knowing that one particle is
on site $\vec 0$ indicated by the cross in the corner of the sample. The grayscale range from $0$ (black) to
$3$ (white).}
\end{figure}
we find that when the FN approximation starts to fail ($r_s\approx 30$), 
the difference between the FN results and the
variational results starts to increase drastically. In addition, the FN results (here for $\rho_2$ and $g_2$)
lies in between the variational and the exact results so that the fixed nodal surface can be viewed as a 
``wall'' preventing the system to relax to the true ground state.
Hence some information on the nature of the ground state can be extracted 
from the evolution of the physical quantities between the variational and the FN estimates (and not only
from the study of the FN quantities alone) This idea will be put to application in 
section~\ref{chara} by looking how the physics put in the various GWF is amplified or washed out
by the FN projection.

\section{Construction of an Hybrid Phase}

The melting point of the Wigner crystal can be viewed as the point above which the problem
is well described in real space (crystal with particles at given positions) and below which
the momentum space is to be used (liquid with well defined Fermi surface). Hence, the melting
point itself is somehow the point of ``maximum uncertainty'' where the problem is equally
badly described in both momentum and real space representation. In the standard scenario of a
(first order) direct transition between the liquid and the crystal, this point is nothing else than
the point where the energy of one state crosses the other one. Here, we consider another scenario which
interpolates between momentum and real space. In order to do so, we introduced in 
Ref.~\onlinecite{falakshahi2005}
a new GWF, called hybrid GWF aimed to provide such an interpolation. As we have seen, we find that this
 hybrid GWF provides the lowest energy in the intermediate region of $r_s$.

The construction of the hybrid GWF $\Psi_{\rm hyb}$ is done in such a way that
the resulting $\phi_i(\vec r_j)$ are the Bloch states of electrons in a triangular crystal.
First, an effective one-body Hamiltonian $H_{\rm eff}$ is constructed for an electron
in a {\it attractive} periodic potential that has the symmetry of the classical (triangular) 
Wigner crystal, 
\be
\label{eq:Heff}
H_{\rm eff}=-t\sum_{\langle\vec r,\vec r'\rangle}c_{\vec r}^\dagger c_{\vec r'}
-U^* \sum_{\vec r} W(\vec r) n_{\vec r} 
\ee
where the one-body potential is $W(\vec r)=\sum_{i=1}^N V(\vec r -\vec u_i)$.
The singularity of $W(\vec r)$ at the position of the classical crystal 
$\vec r=\vec u_i$ has been removed by setting
$W(\vec u_i)\equiv W(\vec u_i +(1,0))$ (we checked that other choices of $W(\vec r)$ give
consistent results). In a second step, $H_{\rm eff}$ is numerically diagonalized using the Lanczos algorithm. 
The $N$ orbitals of lowest energy $\phi_i(\vec r)$ ($1\le i\le N$) are then used to construct the Slater
determinant. {\it A priori} $U^*$ is a variational parameter that allows for an interpolation between
$\Psi_{\rm liq}$ (at $U^*=0$) and localized orbitals leading to a GWF similar to $\Psi_{\rm cry}$ 
(at $U^*\gg 1$).
 
\begin{figure}
\vglue +0.05cm
\includegraphics[width=8cm]{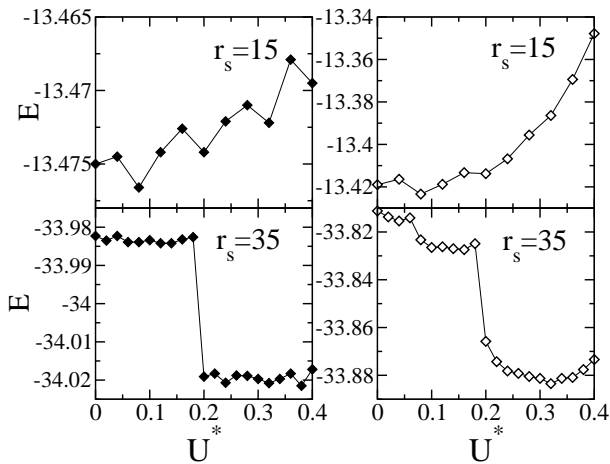}
\caption{\label{fig:dual30} Fixed Node energy (left panels) and variational energy
(right panels) at $r_s=15$ (upper panels) and $r_s=35$ (lower panels) as a function of $U^*$.
 The system contains $30$ particles in $40\times 42$ sites and the calculations are done 
with the hybrid GWF.}
\end{figure}

In Fig.\ref{fig:dual30}, we plot the energy as a function of $U^*$ 
for two values of $r_s$ below ($r_s=15$ upper panels) and above ($r_s=35$ lower panels) 
the expected liquid-crystal transition
(which is found around $r_s\approx 20-25$ for this system of $30$ particles in $40\times 42$ sites).
At $r_s=15$ the energy increases with $U^*$ indicating that the free plane wave solution provides 
the best nodal structure. The increase of the Fixed-Node energy 
(left panels) is however $10$ times smaller than the increase of 
the variational energy and is only slightly above the statistical resolution. At $r_s=35$,
the situation is completely changed, and the FN energy shows a sharp drop around a 
particular critical value $U^*=U^*_C\approx 0.2$. This point corresponds, as we shall see, to
the value of $U^*$ above which the GWF has the symmetry of the crystal.
We find that the fixed node energy is completely flat
above and below this threshold, indicating that the drop in energy is really associated
to the change of symmetry.

\begin{figure}
\vglue +0.05cm
\includegraphics[width=8cm]{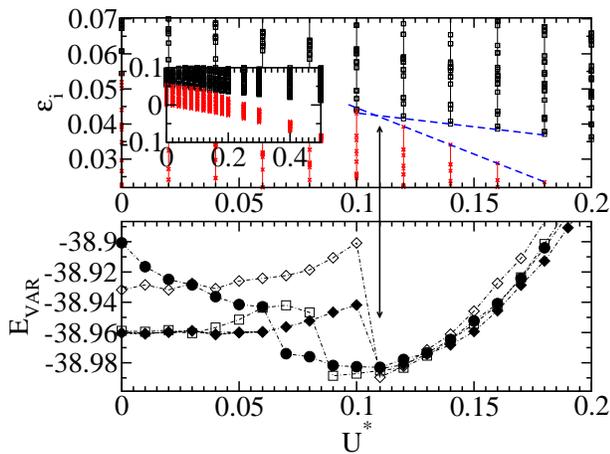}
\caption{\label{fig:band_struct} (color online). Upper panel: lowest energy levels $\epsilon_i$ of the fictitious
one-body problem $H_{\rm eff}$ as a function of the variational parameter $U^*$ for a system of
$72$ particles in $96\times 168$ sites. The dashed lines correspond to the $72^{\rm th}$ and
 $73^{\rm th}$ values of $\epsilon_i$. They indicate the critical value $U^*_c\approx 0.12$ where 
the band of lowest energy splits from the rest of the fictitious spectrum. 
Inset: same thing for a broader range of $U^*$. Lower panel: variational
energy as a function of $U^*$ for $2P^2$ particles in $16P\times 28P$ sites at 
$r_s=40$ with $A=4.5$. $P=3$ (empty diamonds),
$P=4$ (empty squares), $P=5$ (full circles) and $P=6$ (full diamonds).}
\end{figure}

More insight on the significance of the critical value $U^*_C$ can be found by looking 
at the eigenenergies $\epsilon_i$ associated with the orbitals $\phi_i(\vec r )$. The lowest
values of these energies $\epsilon_i$, which have no physical meaning by themselves, are plotted in 
the inset of Fig.~\ref{fig:band_struct} as a function of $U^*$. At $U^*=0$ they form one unique band with
a parabolic dispersion $\epsilon_i \propto k_i^2$. At $U^*\gg 1$, the periodic potential is at the
origin of a more complex band structure, and the band of lowest energy 
(which contains exactly $N$ levels) detaches from the rest of the spectrum and becomes
narrower as $U^*$ increases. What is important here is that the point where this band of lowest energy
separates from the rest of the level (see Fig.\ref{fig:band_struct}, upper paner which is a zoom of the inset)
corresponds exactly to the point where the energy has a drop (see Fig.\ref{fig:band_struct}, lower paner).
Above this point, Bloch theory tells us that the $\phi_i(\vec r )$ can be written in term of the Bloch waves
of one band only:
\be
\label{eq:bloch} 
\phi_i(\vec r )= u_{\vec k_i}(\vec r) e^{i\vec k_i \cdot \vec r}
\ee
where $u_{\vec k_i}(\vec r)$ is a function with the periodicity of the crystal and the momentum
have their value within the first Brillouin zone of the triangular lattice, i.e. an hexagonal Fermi surface.
In that sense, the hybrid GWF is liquid like (made of plane waves) with the triangular symmetry. 
However, since the only thing that enters in the GWF is actually the determinant, 
${\rm det}[ \phi_i(\vec r )]$,
it is equivalent to use (instead of Eq.(\ref{eq:bloch})) linear combinations of the Bloch waves, leading to 
the so called Wannier functions:  
\be
\label{eq:wannier}
\phi_i(\vec r )= F(\vec r -\vec u_j) {\rm \ \ with \ \ }
F(\vec r) \equiv \sum_i u_{\vec k_i}(\vec r) e^{i\vec k_i\cdot\vec r}
\ee
Writing the $\phi_i(\vec r )$ in the form Eq.(\ref{eq:wannier}), the hybrid GWF looks similar 
to the crystal one (for which $F(\vec r)$ takes the form of a gaussian function). 
The Wannier functions are not uniquely defined (since $u_{\vec k_i}(\vec r)$
is defined up to an arbitrary phase) and an important effort has been done in the literature
(in the context of ab initio electronic structure calculation mostly) to define the maximally 
localized Wannier functions. The fondamental result for us is~\cite{kohn1959,blount1962} 
that {\it at $U^*= U^*_C$, $F(\vec r)$ cannot decrease faster than algebrically 
with $\vec r$} and is hence a delocalized function over the sample. This is in sharp contrast 
with the crystal GWF for which the $\phi_i(\vec r )$ are strongly localized around the classical 
positions $\vec u_i$ of the crystal. For $U^*\gg U^*_C$,
the lowest band becomes narrower, and exponentially localized Wannier function $F(\vec r) \sim e^{-h r}$
can be constructed. However, the point of interest for us is  $U^*\approx U^*_C$ where the drop in energy
occurs. In the rest of this paper, we will stick to this value where the Wannier
functions are delocalized and hence completely different from the crystal GWF at a qualitative level.

\section{Characterization of the hybrid phase}
\label{chara}

As we have seen in the introduction (lower panel of Fig.~\ref{fig:intro}), 
the hybrid GWF provides the lowest energy for a rather large domain of $r_s$.
Having established that, we now try to understand the physical origin of this success and the nature
of the corresponding phase. From the construction of the hybrid GWF (Slater determinant
made of a {\it delocalized} Wannier function), we expect (at the variational level) 
something someway between the crystal (it has its triangular symmetry) and the liquid 
(it is made of delocalized one-body functions). While the crystal aspect is straightforward,
the characterization of the liquid aspect is a difficult task within the FN-QMC.
Indeed, the FN-QMC method mainly provides
access to quantities (like $\rho (\vec r)$ or $g (\vec r)$) which are diagonal in the real 
space representation, and hence not very likely to characterize the liquid behaviour.

In Ref.~\onlinecite{falakshahi2005}, we studied some physical quantities for the three GWF with the 
implicit idea that these quantities would characterize the phase associated with each GWF. Here
we follow a different approach. Following the insight we got in section III on a small system,  
we study how the physical quantities evolve between the variational value 
$O_{\rm VAR}=\langle\Psi_G| O|\Psi_G\rangle$ 
and the FN results $O_{\rm FN}=\langle\Psi_0^{\rm FN}| O |\Psi_0^{\rm FN}\rangle$, 
as this gives the general trend toward the true ground state of the system.
The different GWFs capture different physics, which can be either amplified or washed out
by the FN projection procedure, giving hints on the true nature of the ground state.

\begin{figure}
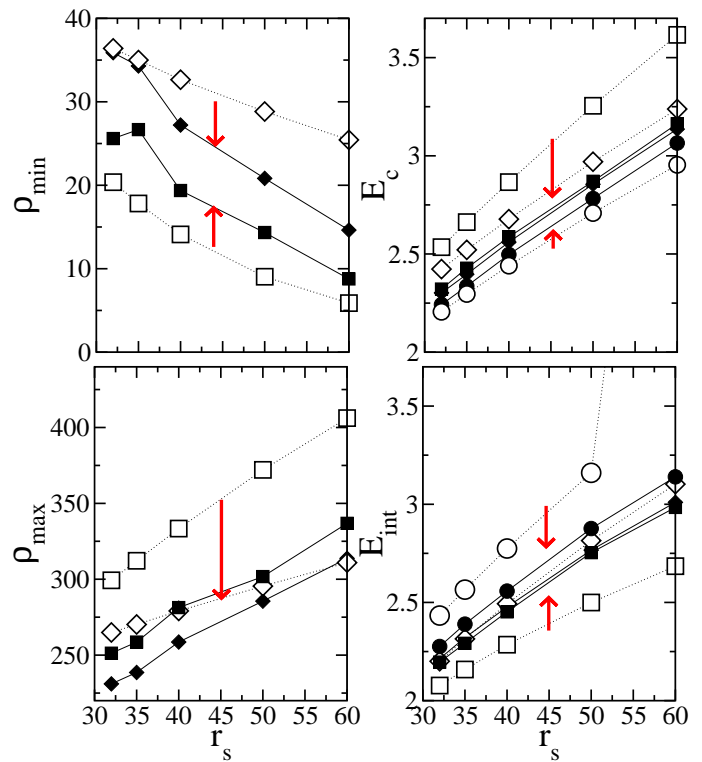

\vglue +0.05cm
\includegraphics[width=9cm]{dst-kin-rs2.eps}
\includegraphics[width=9cm]{dst-kin-rs3.eps}
\caption{\label{fig:dst-kin} Evolution from variational to FN results for various physical
quantities. $\rho_{\rm min}(r_s)=\min_{\vec r} \rho (\vec r)$ (Upper Left panel),
 $\rho_{\rm max}(r_s)=\max_{\vec r} \rho (\vec r)$ (Lower Left panel), $E_c(r_s)$ (Upper Right panel)
and $E_{\rm int}(r_s)$ (Lower Right panel, the Madelung energy $-c_1 r_s/2$ has been substracted)
for a system of $56$ electrons in $56\times 56$ sites. Average done on $5\times  10^5$ walkers. The empty
symbols show the variational results while the full symbols show the FN walking estimates at $\beta=16$
for the hybrid (diamonds), crystal (square) and liquid (circles) GWF. The arrows indicate the evolution
from the variational to the FN estimate. $\rho_{\rm min}$ and $\rho_{\rm max}$ are measured in percentage of the
average density.
}
\end{figure}

\begin{figure}
\vglue +0.05cm
\includegraphics[width=8.1cm]{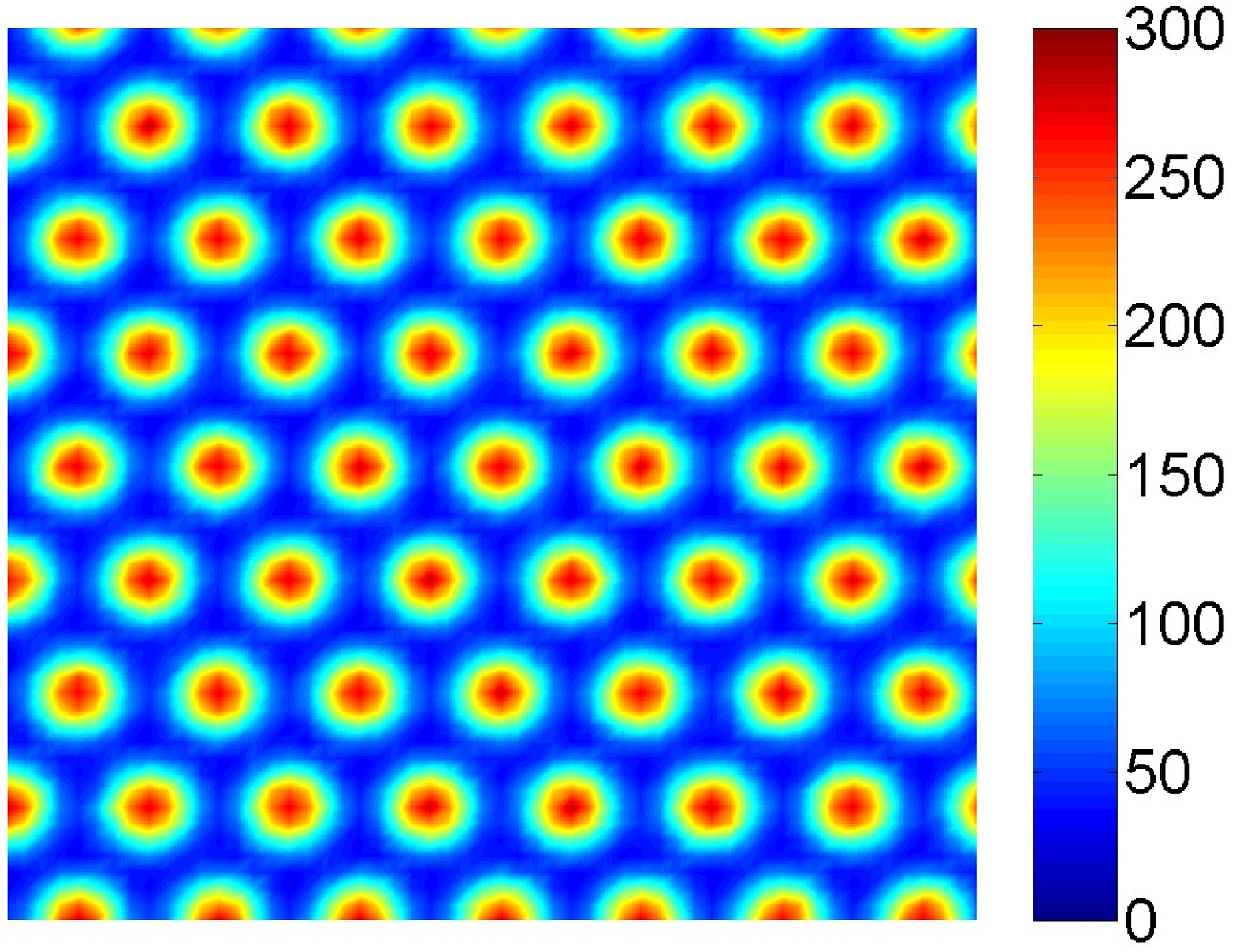}
\includegraphics[width=8cm]{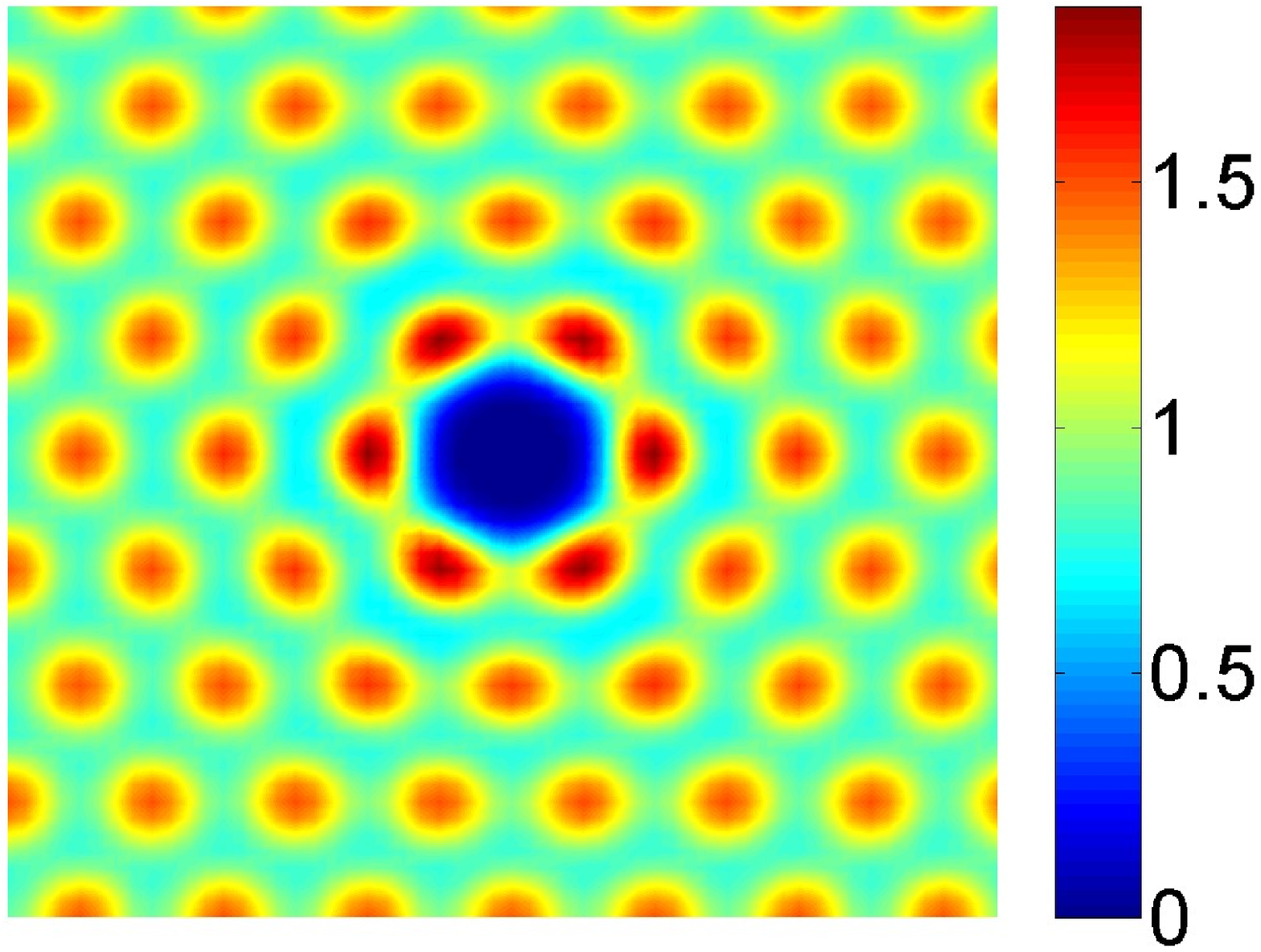}
\caption{\label{fig:3D} (Color online.) Density $\rho(\vec r)$ (Upper panel) and
density-density correlations $g(\vec r)$ (Lower panel) at $r_s=35$
for $56$ electrons in $56\times 56$ for the hybrid GWF. The density is plotted in percentage
of the average density. Its minimum (maximum) correspond to $35\%$ (230\%)
of the average density.
}
\end{figure}

\begin{figure}
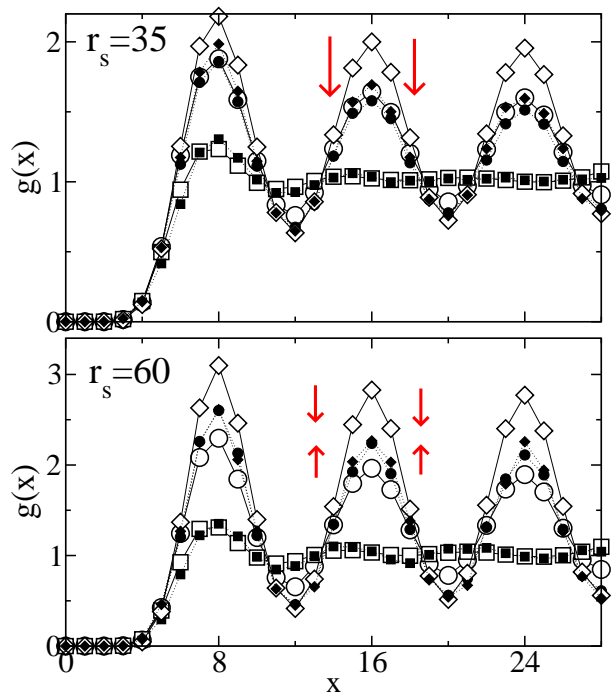

\vglue +0.05cm
\includegraphics[width=8cm]{roro3.eps}
\includegraphics[width=8cm]{roro4.eps}
\caption{\label{fig:roro} Cross section of $g(\vec r)$ with $\vec r=(x,0)$
as a function of $x$  for $56$ electrons in $56\times 56$. The empty (full) symbols are the 
variational (FN) results, respectively for the hybrid (circles), crystal (diamonds)
and liquid (squares) GWF. The upper (lower) panel corresponds to
$r_s=35$ ($r_s=60$). The arrows indicate the direction of the evolution 
from the variational to the FN estimate: while at $r_s=35$ both the crystal and hybrid GWF
tends to get delocalized, at $r_s=60$, the peaks of the hybrid GWF get more pronounced
after the FN projection.
}
\end{figure}

In the right panel of Fig.\ref{fig:dst-kin} we plot the 
kinetic and interaction energies as a function of $r_s$. The variational values of the hybrid GWF
are in the middle of those obtained with the liquid and crystal GWF. Upon applying the FN
projector $e^{-\beta H^{\rm FN}}$ on the hybrid (and crystal) GWF, the value of the kinetic energy
decreases and converges to a value rather close to the variational kinetic energy of the liquid.
It is important to note that these differences in (kinetic or interaction) energy are about 5 to 10
times bigger than the differences observed on the total energy. This decrease of the kinetic energy to an
almost liquid value is a strong sign of the delocalized nature of the ground state. As $r_s$ increased however,
this tendency also diminishes. This delocalization can also be seen indirectly on the electronic density 
$\rho(\vec r)$. An example of $\rho(\vec r)$ is shown in the upper panel of Fig.\ref{fig:3D}. 
At first sight the density looks crystal like (strong peaks forming a triangular lattice). 
However the height of the peaks ($\rho_{\rm max}$ lower left panel of Fig.\ref{fig:dst-kin}) and the depth of the 
valleys ( $\rho_{\rm min}$ upper left panel of Fig.\ref{fig:dst-kin}) show not only that these peaks are not very
pronounced (at $r_s=32$, the density at the peaks is only twice the average density while the ``background''
contains $\rho_{\rm min}\approx 35\%$ of the electrons) but also that the contrast tends to 
diminish from the variational to the FN estimate (especially for the crystal GWF). Similar conclusions 
can be drawn from the density-density correlation function $g(\vec r)$ (A 3D plot is shown in Fig.\ref{fig:3D}. 
Fig.\ref{fig:roro} gives a cross section). 
For $r_s=35$ (upper panel of Fig.\ref{fig:roro}) the peaks of both the hybrid and crystal results tend to
be washed out after the FN projection
At $r_s=60$ however (lower panel of Fig.\ref{fig:roro}) the tendency is inversed and the hybrid result 
tends to get more localized when going from variational to the FN results (the same thing can be seen on
$\rho_{\rm max}$, see lower left panel of Fig.\ref{fig:dst-kin}).

All the physical quantities discussed above indicate the same tendency: 
at the variational level the hybrid GWF lies
somewhere between the liquid and the crystal results, and the FN results get even closer to the liquid 
for not too large $r_s$ (roughly $r_s\le 60$). At a basic level, the raw reason why the hybrid GWF
allows to gain energy with respect to (the previously used) crystal GWF is therefore quite
simple: electrons in the crystal GWF are far too much localized, and the hybrid GWF helps to restore
a better balance between kinetic and electrostatic energy.

\section{Technical aspects}
In this section, we address various technical points. We start with giving some reference values 
of the energy that can be compared directly with the literature. We proceed with studying the
thermodynamic ($N\rightarrow\infty$) and continuum ($\nu\rightarrow 0$) limit. Then, we determine
precisely the critical value $r_s^*\approx 31.5\pm 0.5$ at which the transition between the liquid 
and the hybrid phase takes place. After discussing the status of the mixed estimators, we end the section
with showing that in addition to a upper bound to the energy, a lower bound can also be extracted from
the FN-QMC datas.

\begin{figure}
\vglue +0.05cm
\includegraphics[width=8cm]{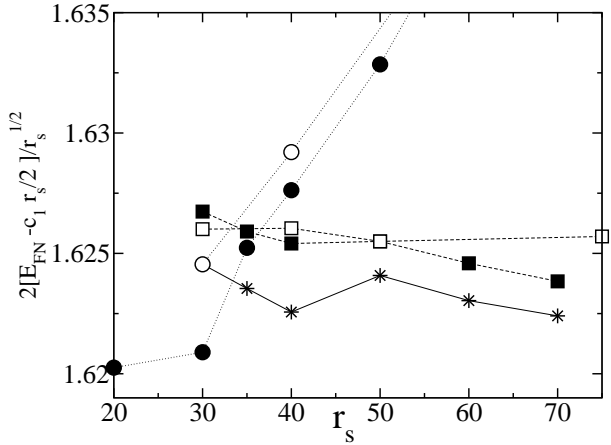}
\caption{\label{fig:energy56vslit} Comparaison with the results of the litterature for a system of
$56$ electrons in $210\times 208$ sites. The plot shows the rescaled energy 
$2 [E_{\rm FN} -c_1 r_s/2]/r_s^{1/2}$ which converges
toward the zero point motion $c_{1/2}$ correction to the energy at high $r_s$. The symbols 
correspond to our calculation with the liquid GWF (full circles), solid GWF (full squares),
hybrid GWF (stars) and to the calculation of Ref.~\onlinecite{rapisarda1996} for the liquid (empty circles, $57$ particles) 
and solid (empty squares).}
\end{figure}

{\it Comparaison with the litterature.}
We start with a discussion of an (almost) square system of $56$ particles in $210\times 208$ 
sites. This system is close to systems previously studied in the litterature, and can hence serve 
as a reference. The absolute values of the FN energies are given in Table.~\ref{table}
together with the results of Ref.~\cite{tanatar1989,rapisarda1996}. The same datas (without 
Ref.~\cite{tanatar1989}) are represented in Fig.~\ref{fig:energy56vslit}
 where we plotted the zero-point motion energy
$2 [E_{\rm FN} -c_1 r_s/2]/r_s^{1/2}$ (which converges toward $c_{1/2}$ at $r_s\gg 1$) as a function of $r_s$.
We find a good agreement with the more recent datas of Ref.~\cite{rapisarda1996} while the original datas
of Ref.\cite{tanatar1989} are about $\sim0.015$ higher (Ref.~\cite{tanatar1989}
and Ref.~\cite{rapisarda1996} used the same numerical code in their calculations). 
The observed small differences can be due to the following systematic errors: (i) too small 
$\beta < \beta_{\rm sat}$. Here we averaged the results over $50\le\beta\le 100$ and some more energy
($~0.003$) can probably be gained by using higher values of $\beta$. Increasing $\beta$ not only
involves more computing time, but also using more walkers to keep the exponential increase of the
variance under control. (ii) Small residual effect of the lattice (see Fig.~\ref{fig:dilu} below), 
(iii) shell effects (Ref.~\cite{tanatar1989,rapisarda1996} used $57$ particles (complete shell) 
for the liquid while we used the same $N=56$ system for all three phases. This effect can account for a 
difference of energy of $\sim 0.004$ at $r_s=40$, see Fig.~\ref{fig:thermo-liq} below), and 
(iv) finite number of walkers.

\begin{widetext}

\begin{table}
\label{table}
\begin{tabular}{|c|c|c|c||c| c |c |c|c|c|c |c |}
\hline
GWF & $N$ & $L_x\times L_y$& &$r_s=20$& $r_s=30$ & $r_s=35$ & $r_s=40$ 
& $r_s=50$ & $r_s=60$ & $r_s=70$ & $r_s=75$ \\
\hline\hline
Liquid & 56 & 210$\times$208&  & -18.499 & -28.744 & -33.906 & -39.097 & -49.532 & -60.017 & & \\
Crystal& 56 & 210$\times$208&  & & -28.728 & -33.904 & -39.104 & -49.558 & -60.074 & -70.634 & \\
Hybrid& 56 & 210$\times$208& $U^*=0.005$ & & -28.734 & -33.911 & -39.113 & -49.563 & -60.080 & -70.640 & \\
\hline
Liquid& 57 & continuum & Ref.~\onlinecite{tanatar1989} & & -28.722 & & -39.073 & & & &-75.788\\
Crystal& 56 & continuum & Ref.~\onlinecite{tanatar1989} & & -28.700 & &-39.090 & -49.526 & & &-75.904 \\
Liquid& 57 & continuum & Ref.~\onlinecite{rapisarda1996} & & -28.734 & & -39.092 & & & &-75.825\\
Crystal& 56 & continuum & Ref.~\onlinecite{rapisarda1996} & & -28.730 & &-39.102 & -49.558 & & &-75.918 \\
\hline
\end{tabular}
\caption{Selected values of energy for the three GWF as well as values from the litterature for comparaison. 
The energies have been averaged over $50/t\le\beta\le 100/t$
and calculated with an average of 25000 walkers. 
The statistical error is of the order of one in the last digit.
The parameter $A$ has been optimized at the variational level
and is very well fitted by $A\approx 0.59\   r_s^{0.57}$ for the liquid and a little lower for the crystal 
while the liquid values have been used for the hybrid GWF. $U^*=0.005$ which corresponds to the splitting of the lowest band. The parameter $d_0$ has also been optimized at the variational level and to very good 
approximation $d_0/a \approx (10/r_s)^{1/4}$.}
\end{table}

\end{widetext}

\begin{figure}
\vglue +0.2cm
\includegraphics[width=8cm]{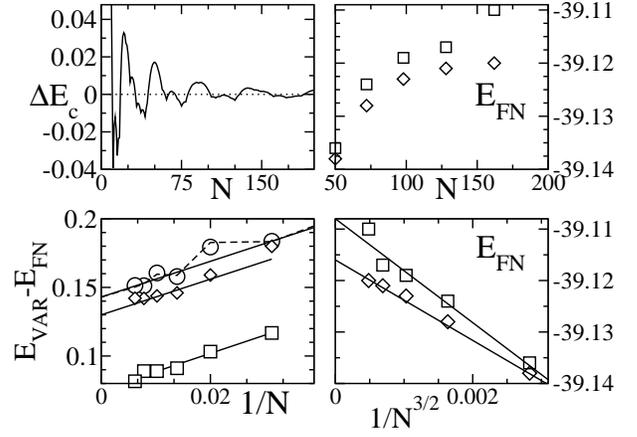}
\caption{\label{fig:thermo-hyb} Finite $N$ effect in a system of 
$N=2P^2$ particles in $16 P\times 28 P$ sites. Upper Left panel: exact finite size correction  
$\Delta E_c (N)$ to the liquid kinetic energy at $r_s=0$. This curve is used to remove the oscillatory
dependance of the liquid energy due to the shell structure. Lower Left panel: difference between the
variational anf fixed-node energy $E_{\rm VAR}-E_{\rm FN}$ as a function of $1/N$ for the liquid (circle,
$A=4.5$), hybrid (diamonds, $U^*=0.12$, $A=4.5$) and crystal (square, $d_0/a=0.74$,$A=4$). The linear fits (solid lines) are respectively $y=0.143+1.3/N$,
 $y=0.13+1.3/N$ and  $y=0.075+1.3/N$. A better fit is obtained for the liquid using (dashed line)
  $y=0.143+1.3/N+0.8\Delta E_c(N)$. Right panels: FN energy for the crystal (squares) and hybrid
 (diamonds, $U^*=0.12$) as a function of $N$ (Upper Right panel) and $1/N^{3/2}$ (Lower Right panel).
The fits (solid lines) are respectively $y=-39.108-10.1/N^{3/2}$ and $y=-39.116-7.8/N^{3/2}$} 
\end{figure}

{\it Thermodynamic ($N\rightarrow\infty$) limit.}
In Fig.~\ref{fig:thermo-liq} we examine the finite $N$ effect for the liquid GWF. Those can be already
understood by looking at the finite $N$ corrections to the energy at $r_s=0$ 
(See Fig.~\ref{fig:thermo-hyb} Upper Left): one finds that the (exact) kinetic energy 
$E_{\rm c}(r_s=0,N)=1+\Delta E_c (N)$ strongly oscillates with $N$ for small values 
of $N$ before saturating toward $1$. These oscillations corresponds to the filling of the various
shells of equal energies. Hence, following Ref.~\cite{tanatar1989}, we fit the $N$ dependance of the 
liquid energy with the form,
\be
E(N) = E_\infty +  E_c(r_s) \Delta E_c (N) - A(r_s)/N
\ee
However, contrary to Ref.~\onlinecite{tanatar1989}, we do not let the coefficient $E_c(r_s)$ to be 
a fitting parameter but rather fix it to the calculated kinetic energy (for a given value of $N$,
the finite $N$ corrections to $E_c(r_s)$ are irrelevant at the precision of the calculations).
Hence, we plot in the inset of Fig.~\ref{fig:thermo-hyb} the liquid energy to which the
$E_c(r_s) \Delta E_c (N)$ correction has been substracted, and find a very good agreement with
a $1/N$ residual error. The quality of the overall (two parameters) fit can be found in the main 
Fig.~\ref{fig:thermo-hyb} for up to $162$ particles. In the Lowest Left panel of Fig.~\ref{fig:thermo-hyb},
we have plotted the difference between the FN and the variational energy as a function of $1/N$
(circles for the liquid). We find that this difference presents some finite $N$ effect about as
strong as those of the FN energy alone (here $E_{\rm FN}\approx -39.088 + 2.5\Delta E_c (N) -1.75/N$
while $E_{\rm VAR}-E_{\rm FN}\approx 0.143+0.8\Delta E_c(N)+1.3/N$.). The difference between the finite $N$ 
behaviour of the variational and FN result is not very surprising since for those intermediates
values of $r_s$ (here $r_s=40$), the difference $E_{\rm VAR}-E_{\rm FN}$ is about ten time bigger 
than those effects. The same conclusion also holds for the hybrid (diamond) and crystal (squares) GWF.
\begin{figure}
\vglue +0.05cm
\includegraphics[width=8cm]{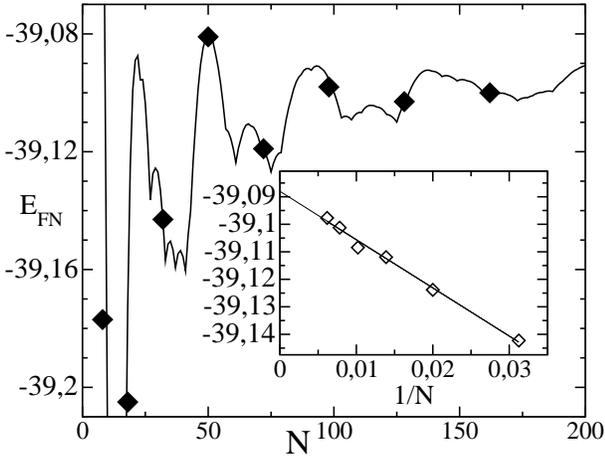}
\caption{\label{fig:thermo-liq} Finite $N$ effect of on the Fixed Node energy with the liquid GWF at $r_s=40$.
Energy averaged over $25/t\le \beta \le 50/t$ as a function of the number $N$ of particles with $A=4.5$
for a system of $N=2P^2$ particles in $16 P\times 28 P$ sites ($P=2\dots 9$ diamonds). The fit (solid line)
corresponds to $E=-39.088 + E_c^{50} (r_s=40) \Delta E_c (N) -1.75/N$ where $E_c^{50} (r_s=40)=2.5$ is the 
(fixed-node) kinetic energy for $50$ particles and $\Delta E_c (N)$ is the exact finite size correction to 
the kinetic energy computed at $r_s=0$. Inset: the oscillatory correction $E_c^{50} (r_s=40) \Delta E_c (N)$
has been substracted from the energy which is plotted as a function of $1/N$. The solid line is a linear
fit to the data $E=-39.088-1.75/N$.}
\end{figure}

The finite $N$ behaviour of the crystal and hybrid results are much simpler. They do not present the
same oscillatory behaviour but rather converge smoothly toward the $N \gg 1$ value (see Upper Right
panel of Fig.~\ref{fig:thermo-hyb}.) A good fit is obtained by plotting the FN energies as a function
of $1/N^{3/2}$ (which is the expected finite effect for the crystal at large $r_s$). We find that the difference between crystal and hybrid slightly increases as $N$ gets bigger (The respective fits are
$E_{\rm FN}=-39.108-10.1/N^{3/2}$ for the crystal GWF and $E_{\rm FN}=-39.116-7.8/N^{3/2}$ for the hybrid.)
We repeated the same procedure at $r_s=50$ (up to $98$ particles only) and arrive at  a similar result
($E_{\rm FN}=-49.56-13.8/N^{3/2}$ for the crystal GWF and $E_{\rm FN}=-49.567-12.2/N^{3/2}$ for the hybrid.)

{\it Continuum ($\nu\rightarrow 0$) limit.} 
The presence of the underlying lattice can induce a correction to the continuum result. 
This effect is a small correction however and should not be mixed up with the much larger effects 
that can take place\cite{falakshahi2004,nemeth2005} at much higher values of $r_s$. In addition, 
contrary to the finite $N$ effect which one would like to avoid 
(since we make use of Ewald resummation, we do not look at mesoscopic
samples in this study), there is an underlying lattice in real (semi-conductor based) samples.
In Fig.\ref{fig:dilu} lower panel, we plot the energy as a function of the inverse of the 
surface of the sample for a fixed number of particle. We find indeed a $1/(L_xL_y)$ correction to the
energy. However (Fig.\ref{fig:dilu} upper panel), this correction is almost GWF independant so
 that the relative stability of the phases is unaffected by the lattice for the $\nu=1/224$ and 
$\nu=1/780$ samples. For the latter, the correction is almost negligeable and
we have a quantitative good agreement with the continuum model 
for the absolute values of the energy (see above, Fig.~\ref{fig:energy56vslit}). 
\begin{figure}
\vglue +0.05cm
\includegraphics[width=8cm]{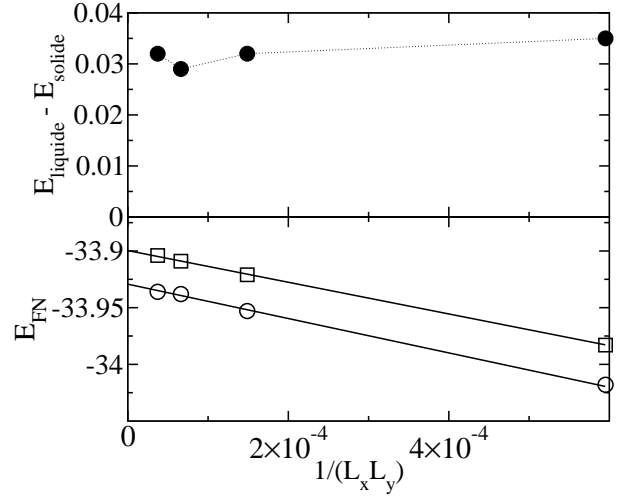}
\caption{\label{fig:dilu} Effect of dilution on the Fixed Node energy for a system of
$30$ particles in $40P\times 42 P$ sites ($P=1,2,3$ and $4$) at $r_s=35$. Lower panel: energy of the liquid GWF
(circles) and solid GWF (squares, $d_0=0.82$) as a function of the inverse volume $1/(L_x L_y)$.
The lines are a guide to the eye. Upper panel: difference of energy between the liquid and the solid GWF
as a function of $1/(L_x L_y)$.}
\end{figure}

\begin{figure}
\vglue +0.05cm
\includegraphics[width=8cm]{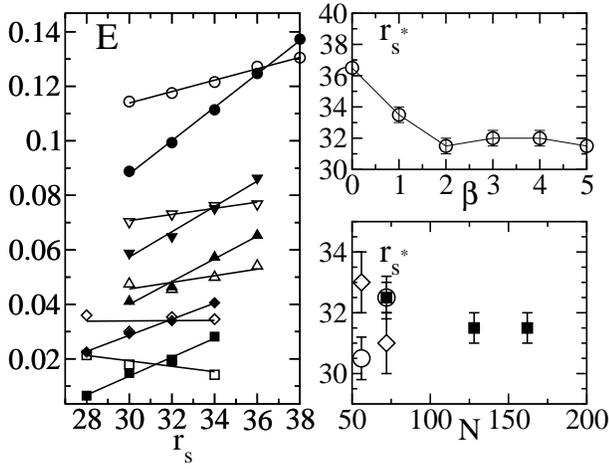}
\caption{\label{fig:rsstar} Determination of the critical value $r_s^*$. Left panel: energy of the liquid (full symbols)
and hybrid (empty symbols) as a function of $r_s$ for $\beta=0,1,2,3$ and $5/t$ 
(top to bottom) with their corresponding linear
fits. The system contains $128$ electrons in $128\times 224$ sites and $-c_1 r_s/2 +c_{1/2} r_s^{1/2}/2$ has been 
substracted to the energy. Upper Right panel: resulting $r_s^*$ (crossing points of the left panel) as a function of $\beta$.
Lower Right panel $r_s^*$ (at $\beta\rightarrow\infty$) as a function of $N$ for a filling factor of
$\nu=1/56$ (empty circles), $\nu=1/224$ (full squares) and $\nu=1/780$ (empty diamonds).
}
\end{figure}

{\it  Critical value of $r_s$.} We now determine precisely the critical value $r_s^*$ at which the 
liquid-hybrid transition takes place (Fig.~\ref{fig:rsstar}). In the left panel of Fig.~\ref{fig:rsstar},
we plot the energy of the hybrid and liquid phase as a function of $r_s$ for various (rather small) values
of $\beta$. For each of them we extract the crossing point $r_s^*(\beta)$ which is plotted in the upper right
panel of Fig.~\ref{fig:rsstar}. We find that $r_s^*(\beta)$ saturates for rather small values 
of $\beta\approx 2/t$ indicating that the gain in energy obtained by 
increasing $\beta$ further ($\beta_{\rm sat}\approx 25/t$ for this system) is almost GWF independant. 
The obtained $r_s^*$ is plotted in the lower part of the right panel as a function of the number 
of particles for several filling factors. At large $N$, $r_s^*$ converges toward $r_s^*= 31.5 \pm 0.5$.

\begin{figure}
\vglue +0.05cm
\includegraphics[width=8cm]{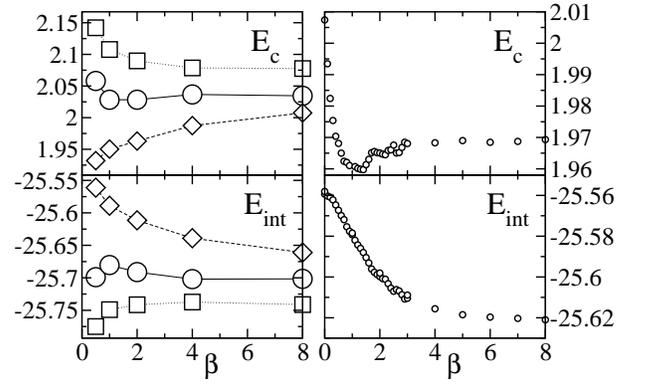}
\caption{\label{fig:Ekinint} Comparaison of the different estimates of the
kinetic and interaction energy for a system of $30$ particles in $40\times 42 $ sites at $r_s=25$.
Left panels: FN estimate for the liquid (diamonds), crystal (squares) and hybrid (circles) GWF. 
Right panels: mixed estimate for the liquid phase. Upper panels: kinetic energy $E_{\rm c}(\beta)$. 
Lower panels: interaction energy $E_{\rm int}(\beta)$}
\end{figure}

{\it Variational, mixed and fixed-node estimates.} 
For a given observable $O$, three different estimates can be constructed. The variational 
$O_{\rm VAR}=\langle\Psi_G| O|\Psi_G\rangle$ and the the fixed node (forward walking) estimates 
$O_{\rm FN}=\langle\Psi_0^{\rm FN}| O |\Psi_0^{\rm FN}\rangle
=\langle\Psi_G|e^{-H^{\rm FN}\beta/2} Oe^{-H^{\rm FN}\beta/2}|\Psi_G\rangle$ have been used through this paper.
An intermediate one, the mixed estimate 
$O_{\rm MX}=\langle\Psi_G| O |\Psi_0^{\rm FN}\rangle=\langle\Psi_G| Oe^{-H^{\rm FN}\beta}|\Psi_G\rangle$ 
is easy to compute and hence very common in QMC calculations. For most applications, $|\Psi_G\rangle$
is a reasonable approximation of the ground state  $|\Psi_0^{\rm FN}\rangle$
so that the mixed results $O_{\rm MX}$ can be extrapolated toward the FN result 
using the (first order correction) formula $O_{\rm FN}\approx 2 O_{\rm MX}- O_{\rm VAR}$. 
The mixed estimate is much easier to obtain than
the FN estimate, since the latter requires to apply $e^{-H^{\rm FN}\beta/2}$ after that
the measurement has been made, and the deaths and births involved in the reconfiguration process 
considerably decrease the statistics. However, in the range of $r_s$ studied in this work,
we find important differences between the different estimates so that it is necessary 
to use the FN estimates. To illustrate this point, we plot in Fig.~\ref{fig:Ekinint}
$O_{\rm FN}(\beta)$ (Left panels) and $O_{\rm MX}(\beta)$ (Right panels) for the kinetic energy
$E_c$ (Upper panels, $E_c\approx 1$ at $r_s=0$) and interaction energy $E_{\rm int}$ (Lower panels).
For the case considered in Fig.~\ref{fig:Ekinint}, the mixed estimate is a very poor estimate of the 
true (FN) estimate. Indeed, we have  $E_c^{\rm VAR}\approx 2.01$, 
$E_c^{\rm MX}(\beta=8)\approx 1.97$ and $E_c^{\rm FN}(\beta=8)\approx 2.01$ so that the interpolation from
the mixed estimate would be very bad.

{\it Lower bound to the total energy. }
The evolution of the kinetic and interaction energy with $\beta$ can be put to further use: we find that
$E_c^{\rm liq}(\beta)< E_c^{\rm hyb}(\beta) < E_c^{\rm cry}(\beta)$ while 
$E_{\rm int}^{\rm cry}(\beta)< E_{\rm int}^{\rm hyb}(\beta) < E_{\rm int}^{\rm liq}(\beta)$
(consistent with the interpretation that the hybrid phase is intermediate between the liquid
and the crystal). As both $E_c^{\rm liq}(\beta)$ and $E_{\rm int}^{\rm cry}(\beta)$ 
increases with $\beta$, (and since by construction the liquid and crystal GWF favor respectively the kinetic 
and interaction energy), we conjecture that they form true {\it lower bounds} of the kinetic and interaction 
energies of the true ground state of the system. Hence, in addition to an upper bound of the total 
energy, we can also construct a lower bound,
\be
E_c^{\rm liq} + E_{\rm int}^{\rm cry} \le E_0 \le E_{\rm FN}^{\rm hyb}
\ee
which for the case of Fig.~\ref{fig:Ekinint} gives $-23.73\le E_0 \le -23.67$ with a precision of $0.3\%$.
This lower bound is of interest for us, as it allows to estimate that the gain in energy
provided by the hybrid GWF (with respect to the crystal and liquid one)  
is a significant fraction of the distance to the lower bond.

\section{Discussion}

We conclude this paper with a discussion of several topics linked with the present study.

{\it Spin of the electrons.} We restricted the present study to fully spin polarized electrons.
Indeed, as the interaction energy gets bigger, the system can minimize its interaction energy
by antisymmetrysing the orbital part of its wave-function, and it was found (for instance in 
Ref.~\cite{rapisarda1996,attaccalite2002}) that above $r_s\ge 20$ (which is the range of our study) 
the polarized liquid is more stable  than the unpolarized one. A strong in-plane magnetic field 
would also polarize the system at a smaller value of $r_s$.

{\it Multiple exchange.} At high $r_s$ when the crystal is well established, multiple 
exchange (of two, three or more particles) can take place~\cite{roger1984} , leading to an effective interaction
between the electronic spins. Recently these exchange interactions were estimated using the
(finite temperature) path integral QMC technique~\cite{bernu2001,bernu2003}. 
The authors proposed that the melting of the crystal is actually due to a divergence of 
multi-spin exchange. In those calculations, some sort of FN approximation
was used below $r_s\approx 60$ to stabilize the crystal. Summing up the exchange energies calculated in
\cite{bernu2003}, we find (in our unit) $E_{\rm ex}\approx 0.001$ at $r_s=50$ (where the calculation is most robust). 
This energy is relatively small and cannot account for the difference observed between the hybrid and 
crystal GWF results. The difference between these path integral calculations and those presented in this paper
lies in the FN approximation (the one used in \cite{bernu2003} seems to stabilize a well localized crystal) and the
use of a rather large temperature $T\approx 0.08$ (which is roughly three times bigger than the energy 
difference between the crystal and the liquid at $r_s=50$ and a 100 times bigger than 
the calculated exchange energies).

{\it Hexatic and supersolid phase.} Among the proposed intermediate phases is the presence of a
hexatic quantum phase in between the liquid and the crystal~\cite{oganesyan2001,barci2003}. Such a phase would
have orientational order, but no translational order. The hybrid GWF is made of Bloch waves that
have, by construction, a hexagonal Fermi surface (one full band corresponds to momentum in the first
Brillouin zone of the triangular crystal). However, we have also explicitely broken the translational symmetry
so that, at the variational level, the hybrid GWF  does certainly not correspond to a hexatic phase. 
Upon applying the FN projection operator, the translational symmetry tend to be restored (see the left panels 
of Fig.~\ref{fig:dst-kin} for instance) so that the present calculations cannot rule out the possibility of 
a quantum hexatic phase. Another proposition is the presence of a supersolid phase~\cite{spivak2003, 
katomeris2003,nemeth2003}, i.e. the quantum coexistence of a liquid (of delocalized defects for instance) 
with the crystal. Qualitatively, the hybrid phase is in agreement with the idea of the supersolid phase as 
discussed in Ref.~\onlinecite{leggett1970}. In the supersolid described in Ref.~\onlinecite{andreev1969} 
however, one expects the number of electrons in the crystal to be smaller than the total number of 
electrons. We tried to construct such a GWF without success.

{\it Bubbles and Stripes. } Around the melting point of a classical liquid-solid phase transition, 
the system gains some energy by being in a coexistence of the two phases. In the present system such
a macroscopic phase separation is not possible since it would lead to a macroscopic dipole. Spivak and Kivelson
recently proposed that a series of microemulsion phases (bubbles/stripes of solid/liquid in the liquid/solid) 
can however take place.~\cite{jamei2005,spivak2004} If they exist, these microemulsions take place between 
the two values of densities $n_s$ and $n_l$ (in electrons per bohr radius squared) 
that come out of the Maxwell construction: $\partial_n E^{\rm sol}(n_s)=
\partial_n E^{\rm liq}(n_l)=[E^{\rm liq}(n_l)-E^{\rm sol}(n_s)]/(n_l-n_s)$
(energies in Rydberg). Solving for $n_s$ and $n_l$ using the FN energies, we find that the (maximum) 
range of $r_s$ where these microemulsions could occur is $|r_s-r_s^*|\le \Delta r_s$ with $\Delta r_s \approx 0.04$ 
($r_s^*$ is the transition point. To a good approximation, $\Delta r_s\approx r_s^* \alpha/c_1$ where the change of slope
$\alpha\approx 0.003$, defined as $E^{\rm liq}\approx E^{\rm sol} + \alpha (r_s -r_s^*)$ around the transition can be extracted
from Fig.\ref{fig:intro} or Fig.\ref{fig:rsstar}.) The argument was originally developped for the liquid/crystal transition but 
holds equally well for the liquid/hybrid transition  described in this paper 
(a direct liquid/crystal transition would lead to $\Delta r_s\approx 0.06$).
The maximum gain of energy given by the Maxwell construction (which uses the energies calculated for {\it neutral} phases, hence only a
fraction of this energy can possibly be gained by microemulsions) is $\Delta E \approx 2\alpha^2/(c_1 r_s^*)\approx 0.2 \mu Ry$  ($\approx 10^{-4}$ in our units).

{\it Metal-Insulator transition.} About ten years ago, a new interest in two dimensional systems arised
from the work of Kravchenko et al~\cite{kravchenko1994} 
who reported on an unexpected metal-insulator transition. 
It was followed by an important body of literature but the origin of this transition is still under 
debate~\cite{abraham2001,kravchenko2004,pudalov2004}. The problem involves understanding the role of disorder in the system and lies outside the scope of this paper. Indeed the effect of disorder is a complicated issue 
and depends not only on the strength of the disorder but also on the typical length scale on which 
the potential varies. For instance, the component of the disorder varying on the scale of the inter electron
distance is likely to pin and stabilize a crystal (or glassy) phase. A disordered potential varying
on larger length scale however will favor fluctuations of the density and hence phase separation. In actual
samples, long length scale fluctuations have been observed using local compressibility 
measurements~\cite{ilani2001}. 

A  natural question for us is weither puddles of 
hybrid/crystal phase are actually present in the experiments. 
As the mobility of the two dimensional systems improved 
in the past years, so had the critical value of $r_s$ where the transition is observed. In hole GaAs 
devices it goes from~\cite{ilani2001,leturcq2003} $r_s=20-25$ to~\cite{noh2003,noh2003b} $r_s\approx 55-60$
in the cleanest samples, i.e. well above $r_s^*=31.5$. At $r_s=60$, the difference of energy between the
liquid and the crystal phase is roughly 0.03 mRy per particle (see lower panel of Fig.~\ref{fig:intro})
so that the melting temperature of the crystal should be slightly larger (the crystal should be 
stabilized by its spin entropy through the Pomeranchuk effect~\cite{spivak2003}) 
which translates into temperatures of a few tens of mK. Those temperatures are therefore within the 
possibilities of a good dilution fridge through slightly below the temperatures that are usually studied.

{\it Conclusion: how liquid is the hybrid phase?}
To summarize the main message of this article, we find that in these intermediate regions
of densities, where little of  the physics is known, the fixed node quantum Monte Carlo technique
should be viewed as a probe. The freedom in the choice of the GWF allows one to  start 
with a GWF that captures some sort of physics (as seen in the variational calculation) and
 the evolution of the various (not only the energy)
physical quantities upon applying the fixed-node projection operator
gives us insights on the physics of the true ground state. In short, one introduces new physics in a GWF
and sees weither this new physics is stabilized or washed out by the FN-QMC algorithm.

With this paradigm in mind, we followed the evolution (from variational to FN) of the physical 
quantities available within our algorithm. For $31.5\le r_s\le 60$, we found that all of them
show the same tendency to delocalization. The succes of the hybrid GWF with respect to the crystal one is 
associated to its (much more) delocalized nature that allow a better balance between kinetic and electrostatic
energy. The success of the hybrid GWF with respect to the liquid one however is associated to the change
in symmetry as the hybrid GWF is constructed out of delocalized waves.
We conclude that while the nature of the ground state above $r_s^*=31.5$ might still not be fully 
elucidated, the fact that it is not a liquid, yet not a localized crystal neither is now put on 
very firm grounds. At large $r_s$ the system eventually gets localized. At the present we cannot say 
weither this occurs through a crossover or a second transition.

{\it Acknowledgment} It is a pleasure to thank
B. Bernu, M. Holzmann, D. L'H\^otes, J. Houdayer, O. Parcollet, F. Portier, J-L. Pichard, P. Roche 
and B. Spivak for very interesting discussions and comments. Special
thanks to Houman Falakshahi with whom this work was initiated.

\newcommand{{{\PRB}}}{{{Phys. Rev. B}}}\newcommand{{{\PRA}}}{{{Phys. Rev. A}}}\newcommand{{{\PRL}}}{{{Phys. Rev. Lett}}}\newcommand{{{\NPB}}}{{{Nucl. Phys.}}}\newcommand{{{\RMP}}}{{{Rev. Mod. Phys.}}}\newcommand{{{\ADV}}}{{{Adv. Phys.}}}\newcommand{{{\EPJB}}}{{{Eur. Phys. J. B}}}

\end{document}